\documentclass[12pt]{article}
\usepackage{graphicx}
\makeatletter
	\renewcommand{\theequation}{
	\thesection./arabic{equation}}
		\@addtoreset{equation}{section}
\makeatother

\newcommand{\mapright}[2]
{\mathop{\hbox to 8mm{\rightarrowfill}}
\limits^{\scriptstyle #1}_{\scriptstyle #2}}

\makeatletter
\@addtoreset{equation}{section}
\makeatother
\renewcommand{\theequation}{\arabic{section}.\arabic{equation}}

\begin{document} %
\begin{flushright}
{KOBE-TH-05-01}\\
\end{flushright}
\vspace{5mm}
\vspace*{6mm}
\begin{center}
{\Large \bf
Supersymmetry in gauge theories \\ with extra dimensions}
\vspace{10mm} \\
C. S. Lim\footnote[1]{e-mail: lim@kobe-u.ac.jp (C.S. Lim)}$^{,a}$,
Tomoaki Nagasawa\footnote[2]{e-mail:
tnagasaw@kobe-u.ac.jp (T. Nagasawa)}$^{,b}$, \\
Makoto Sakamoto\footnote[3]{e-mail:
dragon@kobe-u.ac.jp (M. Sakamoto)}$^{,a}$
and
Hidenori Sonoda\footnote[4]{e-mail:
hsonoda@kobe-u.ac.jp (H. Sonoda)}$^{,a}$
\vspace*{10mm} \\
{\small \it
${}^a $Department of Physics, Kobe University,
Rokkodai, Nada, Kobe 657-8501, Japan
\\
${}^b $Graduate School of Science and Technology, Kobe University,
Rokkodai, Nada, \\ Kobe 657-8501, Japan
\\
}
\vspace{15mm}
{\bf Abstract}
\vspace{4mm}
\\
\begin{minipage}[t]{130mm}
\baselineskip 6mm 

We show that a quantum-mechanical $N=2$ supersymmetry is hidden in 4d mass spectrum 
of any gauge invariant theories with extra dimensions.
The $N=2$ supercharges are explicitly constructed in terms of differential forms.
The analysis can be extended to extra dimensions with boundaries, and for a single extra dimension we clarify a possible set of boundary conditions consistent with 5d gauge invariance, although some of the boundary conditions break 4d gauge symmetries.
\end{minipage}
\end{center}
\vspace{8mm}
\noindent
%
\newpage
\baselineskip 6mm 
\section{Introduction}

Much attention has been paid recently to gauge theories with extra dimensions 
to explore new possibilities for gauge symmetry breaking 
and solving the hierarchy problem 
without introducing additional Higgs fields\cite{
SS,hosotani,G-H,OrbifoldGut,Nagasawa-Sakamoto,Csaki,Higgsless,arkani-cohen-georgi,deconstruction,BRS&Unitarity}.
For instance, in the gauge-Higgs unification scenario, extra components of gauge fields play a role of Higgs fields\cite{hosotani,G-H}.
Attractive models of grand unified theories have been constructed on orbifolds, in which gauge symmetry breaking is caused by orbifolding\cite{OrbifoldGut}.
With extra dimensions with boundaries, Higgsless gauge symmetry breaking can be realized via boundary conditions\cite{Nagasawa-Sakamoto,Csaki,Higgsless,deconstruction}.
The interesting scenario of dimensional 
deconstruction\cite{arkani-cohen-georgi} can be regarded 
as a gauge theory with latticized extra dimension.

In those models, the notorious quadratic divergence problem of scalar fields is absent.
In a higher dimensional point of view, 
this is easily understood because any divergences of mass corrections 
to gauge fields are protected by a higher dimensional gauge invariance.
In a 4-dimensional point of view, however, 
the cancellation of the divergences does not seem to be manifest 
because a remnant of higher dimensional gauge invariance is not apparent 
in 4d effective theories and the cancellation can occur 
only after {\it all} massive Kaluza-Klein modes are taken into account.
(If we truncate massive KK modes at some energy, the cancellation becomes incomplete.)
Furthermore, the cancellation still occurs even when 4d gauge symmetries are broken via orbifolding, the Hosotani mechanism or boundary conditions. 
Thus, in constructing phenomenological models,
it will be important to understand higher-dimensional gauge invariance 
from a 4d effective theory point of view.

Another appealing and well-known scenario to solve the problem of 
the quadratic divergence is to invoke supersymmetry. 
Then, it may be natural to ask a question whether these two 
kinds of theories, gauge theories with extra dimensions and supersymmetric theories, ever have some relation. 
The immediate answer is negative, 
since supersymmetry necessarily needs fermionic degree of freedom, 
while the cancellation mechanism of the quadratic divergence in the higher dimensional gauge theory does not necessitate it. 
We will, however, see that actually these two are related:
(quantum-mechanical) supersymmetry is hidden 
in the higher dimensional gauge theories. 
The main purpose of this paper is to show it.
We note that quantum-mechanical supersymmetry, 
being $0+1$ dimensional field theory, 
can be described without using any spinors.

There exist some evidences for the existence of some kind of supersymmetry 
in 4d mass spectrum already at the truncated low-energy theory, 
as the remnant of higher dimensional gauge symmetry:
If a 4d gauge symmetry is not broken, a massless 4d gauge field appears because 4d gauge invariance guarantees the gauge field to be massless.
This may be explained from a supersymmetry point of view because supersymmetry ensures that the ground state has zero energy and the zero energy state is interpreted as the massless gauge field.
The second evidence is that in a 5d gauge theory with a single extra dimension, every massive mode of $A_{5,n}$ (a massive KK mode of the gauge field in the direction of the extra dimension) can be absorbed into the longitudinal mode of $A_{\mu,n}$ (a massive KK mode of the 4d gauge field) by gauge transformations.
This fact implies that there should exist a one-to-one correspondence between $A_{5,n}$ and $A_{\mu,n}$.
The correspondence may be interpreted as supersymmetry 
between a \lq \lq bosonic" state and a \lq \lq fermionic" one, 
although both modes are bosonic.
The last evidence is that a massless mode $A_{5,0}$ (if exists) cannot be gauged-away 
and it appears as a physical state, in contrast to the massive modes $A_{5,n}$, 
which can be gauged-away and hence are unphysical modes.
This observation is again consistent with supersymmetry 
because zero energy states do not form any supermultiplets 
between bosonic and fermionic states.

In this paper, we show that a quantum-mechanical $N=2$ supersymmetry exists 
in any gauge invariant theories with extra dimensions.
To this end, we construct an $N=2$ superalgebra in terms of differential forms 
in Section 2.
In Section 3, we show that the $N=2$ supersymmetric structure appears 
in 4d mass spectrum of gauge theories with extra dimensions. 
In Section 4, extra dimensions with boundaries are discussed.
For a 4+1-dimensional gauge invariant theory on an interval,
a consistent set of boundary conditions with 5d gauge invariance is 
successfully obtained from a supersymmetric point of view.
The Section 5 is devoted to summary and discussions.
A simple proof of the Hodge decomposition theorem is given in an Appendix.
\section{$N=2$ supersymmetry algebra and differential forms}

In this section, we construct an $N=2$ supersymmetry algebra in terms of differential forms.
We will see later that the $N=2$ supersymmetry is realized in 4d mass spectrum of any gauge invariant theories with extra dimensions.

Let $K$ be a D-dimensional compact Riemannian manifold  
with a metric $g_{ij}\ (i,j =1,2,\cdots,D)$.
A $k$ form $\omega^{(k)}$ on $K$ is given by
\begin{equation}
	\omega^{(k)}=\frac{1}{k!}\omega_{i_1 i_2 \cdots i_k} dy^{i_1} \wedge dy^{i_2} \wedge \cdots \wedge dy^{i_k} ,
\end{equation}
where $y^i\ (i=1,2, \cdots,D)$ are coordinates on $K$ and $\wedge$ denotes the wedge product.
The coefficient $\omega_{i_1 \cdots i_k}$ is totally antisymmetric in all $k$ indices.
The Hodge star (Poincar{\' e} dual) operator on the $k$ form $\omega^{(k)}$ is defined by 
\begin{eqnarray}
	*\omega^{(k)} =\frac{\sqrt{g}}{k! (D-k)!} \omega_{i_1 \cdots i_k} g^{i_1 j_1} \cdots g^{i_k j_k} \epsilon_{j_1\cdots j_k j_{k+1}\cdots j_D} dy^{j_{k+1}} \wedge \cdots \wedge dy^{j_D},
\end{eqnarray}
where $g={\rm det} g_{ij}$ and $\epsilon_{i_1 i_2 \cdots i_D}$ is a totally antisymmetric tensor with $\epsilon_{12\cdots D}=1$.
Repeated applications of $*$ on any $k$ form give
\begin{equation}
	**\omega^{(k)} =(-1)^{k(D-k)}\omega^{(k)}.
\end{equation}
The inner product of any two $k$ forms $\omega^{(k)}$ and $\eta^{(k)}$ is defined by
\begin{eqnarray}
	(\eta^{(k)}, \omega^{(k)})_{\Delta} &\equiv& \frac{1}{k!} \int_{K} d^Dy \ \sqrt{g} \Delta \eta_{i_1 \cdots i_k} \omega_{j_1 \cdots j_k} g^{i_1 j_1}\cdots g^{i_kj_k} \nonumber \\
	&=&\int_{K} \Delta \ \eta^{(k)}\wedge *\omega^{(k)}.
\end{eqnarray}
Here, we have introduced a weight function $\Delta(y)$ for later convenience with a property
\begin{equation}
	\Delta(y) >0.
\end{equation}
Then, it follows that
\begin{eqnarray}
	(\eta^{(k)},\omega^{(k)})_{\Delta} &=&( \omega^{(k)}, \eta^{(k)})_{\Delta}, \\
	(\omega^{(k)},\omega^{(k)})_{\Delta} &\ge& 0, \\
	(\eta^{(k)}, * \omega^{(D-k)})_{\Delta} &=& (-1)^{k(D-k)} (*\eta^{(k)},\omega^{(D-k)})_{\Delta}.
\end{eqnarray}
The adjoint of the exterior derivative $d$ is defined, with respect to the inner product (2.4), by 
\begin{equation}
	(\eta^{(k-1)}, d^{\dagger}\omega^{(k)})_{\Delta} \equiv (d\eta^{(k-1)},\omega^{(k)})_{\Delta}.
\end{equation}
For compact manifold without a boundary\footnote{An extension with boundaries will be discussed in Section 4.}, the action of $d^{\dagger}$ on a $k$ form $\omega^{(k)}$ turns out to be of the form
\begin{equation}
	d^{\dagger}\omega^{(k)} = -(-1)^{(k-1)D} \Delta^{-1} * d \Delta * \omega^{(k)}.
\end{equation}
For $k=0$ and $1$, $d^{\dagger} \omega^{(k)}$ are explicitly written as 
\begin{eqnarray}
	&&d^{\dagger} \omega^{(0)}=0, \\
	&& d^{\dagger} \omega^{(1)}=-\frac{1}{\Delta \sqrt{g}} \partial_j (\Delta \sqrt{g} g^{ij} \omega_i ),
\end{eqnarray}
where $\omega^{(1)} = \omega_{i}dy^i$.
The first equation comes from the fact that $d^{\dagger}$ maps $k$ forms into $k-1$ forms and there is no $-1$ form.
We notice that the nilpotency of $d^{\dagger}$ still holds irrespective of $\Delta(y)$, i.e.
\begin{equation}
	(d^{\dagger})^2 =0.
\end{equation}
 
We can now construct an $N=2$ supersymmetry algebra. 
To this end, we introduce a 2-component vector
\begin{equation}
	| \Omega^{(k)} \rangle \equiv \left( 
		\begin{array}{c}
			\omega^{(k)} \\
			\phi^{(k+1)} 
		\end{array}
	\right),
\end{equation}
where the upper (lower) component consists of a $k$ ($k+1$) form.
The inner product of two 2-component vectors $|\Omega_1^{(k)} \rangle$ and $|\Omega_2^{(k)} \rangle$ is defined by 
\begin{equation}
	\langle \Omega_2^{(k)} | \Omega_1^{(k)}\rangle \equiv (\omega_2^{(k)}, \omega_1^{(k)})_{\Delta} + (\phi_2^{(k+1)}, \phi_1^{(k+1)})_{\Delta}.
\end{equation}
Then, the $N=2$ supercharges $Q_a$ $(a=1,2)$ are given by\footnote{
Here, the inner product (2.4) should be extended for complex forms as
\begin{eqnarray}
	(\eta^{(k)},\omega^{(k)})_{\Delta} \equiv \frac{1}{k!} \int _{K} d^D y \ \sqrt{g} \Delta (\eta_{i_1\cdots i_k})^* \omega_{j_1 \cdots j_k} g^{i_1 j_1}\cdots g^{i_k j_k}, \nonumber 
\end{eqnarray}
and the relation (2.6) is then replaced by
$(\eta^{(k)},\omega^{(k)})_{\Delta} = 
   (( \omega^{(k)}, \eta^{(k)})_{\Delta})^{*}$.
}
\begin{equation}
	Q_1 =\left(
		\begin{array}{cc}
			0 & d^{\dagger} \\
			d & 0 
		\end{array}
	\right),
	\quad 
	Q_2 =\left(
		\begin{array}{cc}
			0 & -id^{\dagger} \\
			id & 0 
		\end{array}
	\right).
\end{equation}
We note that the action of $Q_a$ on $| \Omega^{(k)} \rangle$ is well defined.
It is easy to show that they form the following $N=2$ supersymmetry algebra:
\begin{eqnarray}
	&& \{Q_a, Q_b\} = 2 \delta_{ab} H, \\
	 &&[Q_a , H] = 0, \\
	 &&[(-1)^F, H] = 0, \\
	 &&\{(-1)^F, Q_a \} = 0, \qquad \qquad {\rm for} \ a,b=1,2,
\end{eqnarray}
where the Hamiltonian $H$ and the operator $(-1)^F$ with
$F$ being the \lq \lq fermion" number operator are defined by 
\begin{eqnarray}
	H= \left(	
		\begin{array}{cc}	
			d^{\dagger}d & 0 \\
			0 & d d^{\dagger}
		\end{array}
	\right), \\
	(-1)^F=\left(	
		\begin{array}{cc}	
			1 & 0 \\
			0 & -1
		\end{array}
	\right).
\end{eqnarray}
We may call states with $(-1)^F=+1 \ (-1)$ \lq \lq bosonic" (\lq \lq fermionic") ones.
All the operators $Q_a$, $H$ and $(-1)^F$ are hermitian with respect to the inner product (2.15).

To examine the structure of the $N=2$ supersymmetry algebra, let us consider the following Schr{\" o}dinger-type equations:
\begin{equation}
	 H |\Omega_n^{(k)} \rangle = (m_n^{(k)})^2 |\Omega_n^{(k)} \rangle.
\end{equation}
Since $H= \left(Q_1\right)^2$ and $Q_1^{\dagger} =Q_1$, the eigenvalues $( m_n^{(k)})^2$ are positive semi-definite, i.e.
\begin{equation}
	( m_n^{(k)})^2 \ge 0.
\end{equation}
Since $(-1)^F$ commutes with $H$, we can have simultaneous eigenfunctions of $H$ and $(-1)^F$ such that
\begin{eqnarray}
	H |m_n^{(k)}, \pm \rangle &=& (m_n^{(k)})^2 |m_n^{(k)},\pm \rangle,\\
	(-1)^F |m_n^{(k)}, \pm \rangle &=& \pm |m_n^{(k)},\pm \rangle.
\end{eqnarray}
Since $Q_1$ commutes (anticommutes) with $H\ \left( (-1)^F\right)$, $Q_1 |m_n^{(k)},\pm \rangle$ have the same (opposite) eigenvalues of $H$ $\left( (-1)^F\right)$ as $|m_n^{(k)},\pm \rangle$, i.e.
\begin{eqnarray}
	H \left( Q_1 |m_n^{(k)}, \pm \rangle \right) &=& (m_n^{(k)})^2 \left( Q_1 |m_n^{(k)},\pm \rangle\right),\\
	(-1)^F \left( Q_1 |m_n^{(k)}, \pm \rangle \right) &=& \mp \left( Q_1 |m_n^{(k)},\pm \rangle \right).
\end{eqnarray}
Actually, the states $|m_n^{(k)}, \pm \rangle$ and $Q_1 |m_n^{(k)}, \mp \rangle$ are mutually related, with an appropriate phase convention, as 
\begin{equation}
	Q_1 |m_n^{(k)}, \pm \rangle=m_n^{(k)}  |m_n^{(k)},\mp \rangle.
\end{equation}
Since $|m_n^{(k)}, \pm \rangle$ have the form
\begin{eqnarray}
	|m_n^{(k)}, + \rangle &=&\left( \begin{array}{c} \omega_n^{(k)} \\0 \end{array} \right), \\
	|m_n^{(k)}, - \rangle &=&\left( \begin{array}{c} 0\\ \phi_n^{(k+1)}\end{array} \right),
\end{eqnarray}
the relations (2.29) are rewritten as 
\begin{eqnarray}
	d\omega_n^{(k)} &=&m_n^{(k)} \phi_n^{(k+1)}, \\
	d^{\dagger}\phi_n^{(k+1)} &=& m_n^{(k)} \omega_n^{(k)}.
\end{eqnarray}
Therefore, there is a one-to-one correspondence between the eigenstates of $d^{\dagger}d$ and $d d^{\dagger}$ for $k$ and $k+1$ forms, respectively, and the eigenvalues are, in general, doubly degenerate except for $m_n^{(k)}=0$.

In Appendix, we have shown that any $k$ form $A^{(k)}$ can be expanded as
\begin{eqnarray}
	A^{(k)} =\sum_{p=1}^{b_k} c_p \eta_p^{(k)} +\textstyle{\displaystyle{\sum_{n_k}}'} a_{n_k} \omega_{n_k}^{(k)}+\textstyle{\displaystyle{\sum_{n_{k-1}}}'}  b_{n_{k-1}} \phi_{n_{k-1}}^{(k)},
\end{eqnarray}
where $\{ \eta_p^{(k)}, p=1,2,\cdots, b_k \}$ is a complete set of the harmonic $k$ forms and $\omega_{n_k}^{(k)}$ and $\phi_{n_{k-1}}^{(k)}$ are eigenfunctions of the equations
\begin{eqnarray}
	d^{\dagger}d \omega_{n_k}^{(k)}&=& (m_{n_k}^{(k)} )^2 \omega_{n_k}^{(k)} \qquad \qquad \quad  \quad \ {\rm for} \  m_{n_k}^{(k)} \ne 0, \\
	 dd^{\dagger} \phi_{n_{k-1}}^{(k)}&=& (m_{n_{k-1}}^{(k)} )^2 \phi_{n_(k-1)}^{(k)} \qquad \qquad \ {\rm for} \  m_{n_{k-1}}^{(k)} \ne 0.
\end{eqnarray}
The summations in eq. (2.34) should be taken over the eigenfunctions with nonzero eigenvalues.
Thus, we have found the following structure among a sequence of differential forms:
\setlength{\unitlength}{1mm}
	\begin{picture}(200,90)(-5,0)
		\put(25,0){\makebox(10,10)[c]{$\vdots$}}
		\put(25,5){\makebox(10,10)[c]{$k-1$ form}}
		\put(25,25){\makebox(10,10)[c]{$k$ form}}
		\put(25,45){\makebox(10,10)[c]{$k+1$ form}}
		\put(25,65){\makebox(10,10)[c]{$k+2$ form}}
		\put(25,75){\makebox(10,10)[c]{$\vdots$}}
		\put(55,50){\makebox(10,20)[c]{$0$}}
		\put(60,53){\vector(0,1){5}}
		\put(48,50){\makebox(10,10)[r]{$d$}}
		\put(55,45){\makebox(10,10)[c]{$(\ \eta^{(k+1)},$}}
		\put(60,47){\vector(0,-1){5}}
		\put(62,40){\makebox(10,10)[l]{$d^{\dagger}$}}
		\put(55,30){\makebox(10,20)[c]{$0$}}
		\put(65,75){\makebox(10,10)[c]{$\ddots$}}
		\put(75,65){\makebox(10,10)[c]{$\phi^{(k+2)}$}}
		\put(79,53){\vector(0,1){15}}
		\put(81,68){\vector(0,-1){15}}
		\put(83,50){\makebox(10,20)[l]{$d^{\dagger}$}}
		\put(67,50){\makebox(10,20)[r]{$d$}}
		\put(75,45){\makebox(10,10)[c]{$\omega^{(k+1)},$}}
		\put(80,47){\vector(0,-1){5}}
		\put(82,40){\makebox(10,10)[l]{$d^{\dagger}$}}
		\put(75,30){\makebox(10,20)[c]{$0$}}
		\put(80,33){\vector(0,1){5}}
		\put(68,30){\makebox(10,10)[r]{$d$}}
		\put(75,25){\makebox(10,10)[c]{$(\ \eta^{(k)},$}}
		\put(80,47){\vector(0,-1){5}}
		\put(80,27){\vector(0,-1){5}}
		\put(82,20){\makebox(10,10)[l]{$d^{\dagger}$}}
		\put(75,10){\makebox(10,20)[c]{$0$}}
		\put(95,45){\makebox(10,10)[c]{$\phi^{(k+1)}\ )$}}
		\put(100,53){\vector(0,1){5}}
		\put(88,50){\makebox(10,10)[r]{$d$}}
		\put(95,50){\makebox(10,20)[c]{$0$}}
		\put(99,33){\vector(0,1){15}}
		\put(101,48){\vector(0,-1){15}}
		\put(103,30){\makebox(10,20)[l]{$d^{\dagger}$}}
		\put(87,30){\makebox(10,20)[r]{$d$}}
		\put(95,25){\makebox(10,10)[c]{$\omega^{(k)},$}}
		\put(100,27){\vector(0,-1){5}}
		\put(102,20){\makebox(10,10)[l]{$d^{\dagger}$}}
		\put(95,10){\makebox(10,20)[c]{$0$}}
		\put(115,30){\makebox(10,20)[c]{$0$}}
		\put(120,33){\vector(0,1){5}}
		\put(108,30){\makebox(10,10)[r]{$d$}}
		\put(115,25){\makebox(10,10)[c]{$\phi^{(k)}\ )$}}
		\put(119,13){\vector(0,1){15}}
		\put(121,28){\vector(0,-1){15}}
		\put(123,10){\makebox(10,20)[l]{$d^{\dagger}$}}
		\put(107,10){\makebox(10,20)[r]{$d$}}
		\put(115,5){\makebox(10,10)[c]{$\omega^{(k-1)}$}}
		\put(125,0){\makebox(10,10)[c]{$\ddots$}}
	\end{picture}
\section{Supersymmetry in gauge theories with extra dimensions}

In this section, with the help of the previous analysis, we show that an $N=2$ supersymmetry is hidden in 4d mass spectrum of any gauge invariant theories with compact extra dimensions without a boundary. 
To this end, we consider a (4+D)-dimensional abelian gauge theory with a weight function.
The (4+D)-dimensional metric is assumed to be of the form
\begin{eqnarray}
	d\tilde{s}^2 = e^{-\frac{4}{D} W(y)} \left(
		\eta_{\mu \nu} dx^{\mu}dx^{\nu} +g_{ij}(y) dy^i dy^j \right).
\end{eqnarray}
The (4+D)-dimensional coordinates are denoted by $x^M = (x^{\mu}, y^i)$, where $x^{\mu}\ (\mu =0,1,2,3)$ are the 4-dimensional coordinates and $y^i\ (i=1,2,\cdots,D)$ are the extra D-dimensional coordinates.
The $\eta_{\mu \nu}$ is a 4d Minkowski metric with $\eta_{\mu \nu}= {\rm diag} (-1,1,1,1)$, and $W(y)$ and $g_{ij}(y)$ are assumed to depend only on the coordinates $y^i$.
We note that for $D=1$ and $W(y)=\frac{1}{2} k|y|$ with $g_{55}(y)=e^{4W(y)}$, the metric reduces to the warped metric discussed by Randall and Sundrum\cite{R-S}.
The action we consider is 
\begin{eqnarray}
	S &=& \int d^4 x {\cal L}_{K} \nonumber \\
	&=& \int d^4 x \int d^D y \sqrt{-G} \widetilde{\Delta} \left\{	-\frac{1}{4} G^{MM'}G^{NN'}F_{MN}F_{M'N'} \right\},
\end{eqnarray}
where $\widetilde{\Delta}(y)$ is a weight function depending on $y^i$ and 
\begin{eqnarray}
	&&F_{MN}(x,y) =\partial_M A_N(x,y) -\partial_N A_M(x,y) ,\\
	&& G_{MN}(y) =\left(
		\begin{array}{cc}
			e^{-\frac{4}{D} W(y)} \eta_{\mu \nu} &0\\
			0& e^{-\frac{4}{D} W(y)} g_{ij}(y) 
		\end{array}
	\right), \\
	&& G(y) = {\rm det} G_{MN}(y).
\end{eqnarray}
For our purpose, it is convenient to rewrite ${\cal L}_{K}$ into the form
\begin{equation}
	{\cal L}_K =\int d^D y \sqrt{g} \Delta \left\{
		-\frac{1}{4} g^{MM'}g^{NN'}F_{MN} F_{M'N'} \right\},
\end{equation}
where
\begin{eqnarray}
	&&g_{MN}(y) = \left(
		\begin{array}{cc}
			\eta_{\mu \nu} &0 \\
			0 & g_{ij}(y) 
		\end{array}
	\right), \\
	&& g(y) ={\rm det} g_{ij}(y) , \\
	&&\Delta(y) =\widetilde{\Delta}(y) e^{-2W(y)}.
\end{eqnarray}
Thus, the system becomes identical to that of the gauge theory with the metric 
\begin{equation}
	ds^2 =\eta_{\mu \nu}dx^{\mu}dx^{\nu} +g_{ij}(y) dy^i dy^j 
\end{equation}
and the weight function $\Delta(y)$.

In a viewpoint of the extra D-dimensions, the 4-dimensional gauge fields $A_{\mu} (x,y)$ are regarded as $0$ forms, while the extra D-dimensional components $A_i(x,y)$ are regarded as a $1$ form, so that we may write
\begin{equation}
	A^{(1)} (x,y) \equiv A_i (x,y) dy^i.
\end{equation}
Here, the exterior derivative $d$ is defined by $d = dy^i \partial_i$ with respect to the coordinates on the extra dimensions.
As was done in the previous section, we introduce the inner product for $k$ forms as
\begin{eqnarray}
	(\eta^{(k)}, \omega^{(k)})_{\Delta} &=& \frac{1}{k!} \int d^Dy \ \sqrt{g(y)} \Delta(y) \eta_{i_1 \cdots i_k}(y) 
	\omega_{j_1 \cdots j_k}(y) g^{i_1 j_1}(y)\cdots g^{i_kj_k}(y).
\end{eqnarray}
It turns out that ${\cal L}_{K}$ can be written into the form
\begin{eqnarray}
	{\cal L}_{K} &=& -\frac{1}{4}\left( \partial_{\mu}A_{\nu}-\partial_{\nu}A_{\mu}, \partial^{\mu}A^{\nu} -\partial^{\nu}A^{\mu} \right)_{\Delta} \nonumber \\
	&& -\frac{1}{2}\left( dA_{\mu}-\partial_{\mu}A^{(1)},dA^{\mu} -\partial^{\mu}A^{(1)}\right)_{\Delta} \nonumber \\
	&&-\frac{1}{2}\left( dA^{(1)},dA^{(1)}\right)_{\Delta}.
\end{eqnarray}
As proved in Appendix, $0$ forms $A_{\mu}(x,y)$ can be expanded as\footnote{We have used the fact that the $0$th Betti number $b_0$ is equal to $1$.}
\begin{eqnarray}
	A_{\mu}(x,y)&=& A_{\mu,0}(x)\eta^{(0)}+\textstyle{\displaystyle{\sum_{n_0}}'} A_{\mu,n_0}(x) \omega_{n_0}^{(0)}(y) \nonumber \\
	&=& A_{\mu,0}(x) \eta^{(0)} +\textstyle{\displaystyle{\sum_{n_0}}'} \displaystyle{\frac{A_{\mu,n_0}(x)}{m_{n_0}^{(0)}}}d^{\dagger}\phi_{n_0}^{(1)}(y) ,
\end{eqnarray}
where
\begin{eqnarray}
	\eta^{(0)} &=&\left( \int d^D y \sqrt{g} \Delta \right)^{-1/2} ,\\
	d^{\dagger}d \omega_{n_0}^{(0)}&=&(m_{n_0}^{(0)})^2 \omega_{n_0}^{(0)} ,\\
	 d d^{\dagger}\phi_{n_0}^{(1)}&=&( m_{n_0}^{(0)})^2 \phi_{n_{0}}^{(1)}, \qquad \qquad {\rm for} \ m_{n_0}^{(0)} \ne 0 , 
\end{eqnarray}
and 
\begin{equation}
	\omega_{n_0}^{(0)} =\frac{1}{m_{n_0}^{(0)}}d^{\dagger} \phi_{n_0}^{(1)} \qquad {\rm or}\qquad \phi_{n_0}^{(1)} =\frac{1}{m_{n_0}^{(0)}}d\omega_{n_0}^{(0)}.
\end{equation}
Since $A^{(1)}(x,y) =A_i (x,y) dy^i$ is a $1$ form, it can be expanded as
\begin{eqnarray}
	A^{(1)}(x,y) &=& \sum_{p=1}^{b_1}\varphi_p(x) \eta_p^{(1)}(y) +\textstyle{\displaystyle{\sum_{n_1}}'} \Phi_{n_1}(x) \omega_{n_1}^{(1)}(y)+\textstyle{\displaystyle{\sum_{n_0}}'} h_{n_0}(x) \phi_{n_0}^{(1)}(y) \nonumber \\
	&=& \sum_{p=1}^{b_1} \varphi_p(x) \eta_p^{(1)}(y) +\textstyle{\displaystyle{\sum_{n_1}}'} \displaystyle{\frac{\Phi_{n_1}(x)}{m_{n_1}^{(1)}}} d^{\dagger} \phi_{n_1}^{(2)}(y) +\textstyle{\displaystyle{\sum_{n_0}}'}\displaystyle{\frac{h_{n_0}(x)}{m_{n_0}^{(0)}}}d\omega_{n_0}^{(0)}(y),
\end{eqnarray}
where $b_1$ is the 1st Betti number and 
\begin{eqnarray}
	&&d\eta_p^{(1)}=0=d^{\dagger}\eta_p^{(1)}, \qquad \qquad \qquad p=1,2,\cdots,b_1, \\
	&&d^{\dagger}d \omega_{n_1}^{(1)} = ( m_{n_1}^{(1)})^2 \omega_{n_1}^{(1)}, \\
	&&dd^{\dagger} \phi_{n_1}^{(2)} =(m_{n_1}^{(1)})^2 \phi_{n_1}^{(2)}, \qquad \qquad {\rm for}  \ m_{n_1}^{(1)} \ne 0 ,
\end{eqnarray}
and 
\begin{equation}
	\omega_{n_1}^{(1)}=\frac{1}{m_{n_1}^{(1)}}d^{\dagger}\phi_{n_1}^{(2)} \qquad {\rm for} \qquad \phi_{n_1}^{(2)} =\frac{1}{m_{n_1}^{(1)}}d\omega_{n_1}^{(1)}.
\end{equation}
We should notice that the modes $\omega_{n_0}^{(0)}$ in $A_{\mu}(x,y)$ and $\phi_{n_0}^{(1)}$ in $A^{(1)}(x,y)$ form a supermultiplet, as shown in the previous section.
Inserting the mode expansions (3.14) and (3.19) into the Lagrangian ${\cal L}_{K}$ and using the orthogonal relations of the eigenfunctions, we have
\begin{eqnarray}
	{\cal L}_{K} &=&-\frac{1}{4} \left(F_{\mu\nu,0}(x) \right)^2 \nonumber \\
	&& +\textstyle{\displaystyle{\sum_{n_0}}'} \displaystyle{\left[-\frac{1}{4}\left(F_{\mu\nu,n_{0}}(x)\right)^2-\frac{(m_{n_0}^{(0)})^2}{2}\left( A_{\mu,n_0}(x)-\frac{1}{m_{n_0}^{(0)}}\partial_{\mu} h_{n_0}(x)\right)^2 \right] }\nonumber \\
	&& -\frac{1}{2} \sum_{p=1}^{b_1} \left(\partial_{\mu} \varphi_p (x)\right)^2 \nonumber \\
	&& +\textstyle{\displaystyle{\sum_{n_1}}'} \displaystyle{\left[-\frac{1}{2}\left(\partial_{\mu}\Phi_{n_1}(x) \right)^2 -\frac{(m_{n_1}^{(1)})^2}{2}\left( \Phi_{n_1}(x)\right)^2 \right]},
\end{eqnarray}
where
\begin{equation}
	F_{\mu\nu,n}(x) =\partial_{\mu}A_{\nu,n}(x) -\partial_{\nu} A_{\mu,n}(x).
\end{equation}
Therefore, we conclude that in a 4-dimensional point of view, the field contents of the model are given as follows:
$A_{\mu,0}$ is a massless gauge field.
$A_{\mu,n_0}$ are massive vector bosons with mass $m_{n_0}^{(0)}$.
$h_{n_0}$ are would-be Nambu-Goldstone bosons and can be absorbed into the longitudinal modes of $A_{\mu,n_0}$.
$\varphi_p\ (p=1,2,\cdots,b_1)$ are massless scalars and cannot be gauged-away.
They could play a role of Higgs fields for non-abelian gauge theories\cite{hosotani}.
$\Phi_{n_1}$ are massive scalars with mass $m_{n_1}^{(1)}$.
The origin of the scalar fields $\varphi_p$ and $\Phi_{n_1}$ are the extra dimensional components of the gauge fields.
\section{Extra dimensions with boundaries}

In this section, we extend the previous analysis to extra dimensions with boundaries.
In this case, we have to impose boundary conditions at the boundaries.
The criteria of obtaining a possible set of boundary conditions are, however, less obvious.
For instance, the Dirichlet boundary conditions are used in Higgsless gauge symmetry breaking scenario\cite{Nagasawa-Sakamoto,Csaki,Higgsless}, but the boundary conditions break 4d gauge symmetries explicitly.
Thus, it is not clear whether such boundary conditions lead to consistent gauge theories.
Recently, a criterion to select a possible set of boundary conditions has been proposed in ref. \cite{Csaki}. 
The authors require the boundary conditions to obey the least action principle.
Since the requirement does not, however, rely on gauge invariance directly, it is still unclear that such boundary conditions lead to consistent gauge theories.
Since gauge symmetry breaking can occur via boundary conditions, it is important to clarify a class of boundary conditions compatible with higher-dimensional gauge invariance.

In the following, we discuss how to obtain a possible set of boundary conditions compatible with gauge invariance from a supersymmetry point of view.
To this end, let us consider a 4+1-dimensional abelian gauge theory on an interval
\begin{eqnarray}
	S=\int d^4 x \int_0^L dy \sqrt{-g(y)} \left\{ -\frac{1}{4} F_{MN}(x,y)F^{MN}(x,y) \right\}
\end{eqnarray}
with a non-factorizable metric
\begin{equation}
	ds^2 =e^{-4W(y)}\eta_{\mu \nu}dx^{\mu}dx^{\nu} +g_{55}(y)dy^2.
\end{equation}
The metric reduces to the warped metric discussed by Randall and Sundrum\cite{R-S} when $g_{55}(y)=1$ and $W(y)=\frac{1}{2}k|y|$.
Another choice of $g_{55}(y)=e^{-4W(y)}$ leads to the model discussed in ref. \cite{Nagasawa-Sakamoto}, in which a hierarchical mass spectrum has been observed.

In order to expand the 5d gauge fields $A_{\mu}(x,y)$ and $A_5(x,y)$ into mass eigenstates, we follow the discussions in Section 3 and consider the supersymmetric Hamiltonian,\footnote{
Here, we have represented $H$ and $Q$ in terms of the differential
operator $\partial_y$, instead of $d$ and $d^{\dagger}$, so that 
they act on functions rather than forms.
}
\begin{eqnarray}
	H&=&Q^2 =\left(
		\begin{array}{cc}
			-\frac{1}{\sqrt{g_{55}}}\partial_y \frac{e^{-4W}}{\sqrt{g_{55}}}\partial_y &0 \\
			0& -\partial_y\frac{1}{\sqrt{g_{55}}}\partial_y \frac{e^{-4W}}{\sqrt{g_{55}}}
		\end{array}
	\right), \\
	 Q&=&\left(
		\begin{array}{cc}
			0 & -\frac{1}{\sqrt{g_{55}}}\partial_y \frac{e^{-4W}}{\sqrt{g_{55}}} \\
			\partial_y & 0
		\end{array}
	\right)
\end{eqnarray}
which act on two-component vectors
\begin{equation}
	|\Psi \rangle =\left(
		\begin{array}{c}	
			f(y) \\
			g(y)
		\end{array}
	\right).
\end{equation}
The inner product of two states $|\Psi_1\rangle$ and $|\Psi_2 \rangle$ is defined by 
\begin{eqnarray}
	\langle \Psi_2 | \Psi_1 \rangle =\int_0^L dy \sqrt{g_{55}(y)} \left\{
		f_2(y) f_1(y) + \frac{e^{-4W(y)}}{g_{55}(y)}g_2(y)g_1(y) \right\}.
\end{eqnarray}
To obtain consistent boundary conditions for the functions $f(y)$ and $g(y)$ in $|\Psi\rangle$, we first require that the supercharge $Q$ is hermitian with respect to the inner product (4.6), i.e.
\begin{equation}
	\langle \Psi_2 | Q\Psi_1 \rangle =\langle Q \Psi_2 | \Psi_1 \rangle.
\end{equation}
It turns out that the functions $f(y)$ and $g(y)$ have to obey one of the following four types of boundary conditions\footnote{If we allow $f(L)$ $(g(L))$ to be connected with $f(0)$ $(g(0))$, we have a one parameter family of the boundary conditions\cite{Nagasawa-Sakamoto-Takenaga}:$$\sin{\theta}\ f(0) -\cos{\theta}\ f(L)=\cos{\theta} \ \left(\frac{e^{-4W}}{\sqrt{g_{55}}}g\right)(0)-\sin{\theta}\ \left(\frac{e^{-4W}}{\sqrt{g_{55}}}g\right)(L)=0.$$}:
\begin{eqnarray}
	& {\rm i)} & g(0) =g(L) =0 , \\
	& {\rm ii)} & f(0)=f(L)=0,\\
	& {\rm iii)} & g(0)=f(L)=0, \\
	& {\rm iv)} & f(0)=g(L)=0. 
\end{eqnarray}
We further require that the state $Q |\Psi\rangle$ obeys the same boundary conditions as $|\Psi\rangle$, otherwise $Q$ is not a well defined operator and \lq\lq bosonic" and \lq\lq fermionic" states would not form supermultiplets.
The requirement leads to 
\begin{eqnarray}
	&& \partial_y f(0) =\partial_y f(L) =0 \qquad \qquad \qquad \qquad \qquad \  {\rm for}\ {\rm i)},\\
	&& \partial_y \left( \frac{e^{-4W}}{\sqrt{g_{55}}} g \right)(0) =\partial_y \left( \frac{e^{-4W}}{\sqrt{g_{55}}} g \right)(L) =0 \qquad \ {\rm for} \ {\rm ii)}, \\
	&&\partial_y f(0)=\partial_y \left( \frac{e^{-4W}}{\sqrt{g_{55}}} g \right)(L)=0 \qquad \qquad \qquad \  {\rm for}\ {\rm iii)}, \\
	&&\partial_y  \left( \frac{e^{-4W}}{\sqrt{g_{55}}} g \right)(0) =\partial_y f(L)=0 \qquad \qquad \qquad \  {\rm for}  \ {\rm iv)}.
\end{eqnarray}
Combining the conditions (4.8)-(4.11) together with (4.12)-(4.15), we have found the four types of boundary conditions compatible with supersymmetry,
\begin{eqnarray}
	&&{\rm Type}\ ({\rm N,N}) : \left\{	
		\begin{array}{l}
			\partial_y f(0) =\partial_y f(L) =0 , \\
			g(0) =g(L) =0,
		\end{array}
	\right. \\
	&&{\rm Type}\ ({\rm D,D}) : \left\{	
		\begin{array}{l}
			f(0)=f(L)=0, \\
			\partial_y \left( \frac{e^{-4W}}{\sqrt{g_{55}}} g \right)(0)=\partial_y \left( \frac{e^{-4W}}{\sqrt{g_{55}}} g \right)(L)=0,
			
		\end{array}
	\right. \\
	&&{\rm Type}\ ({\rm N,D}) : \left\{	
		\begin{array}{l}
			\partial_y f(0) = f(L) =0 , \\
			g(0) =\partial_y \left( \frac{e^{-4W}}{\sqrt{g_{55}}} g \right)(L)=0,
		\end{array}
	\right. \\
	&&{\rm Type}\ ({\rm D,N}) : \left\{	
		\begin{array}{l}
			f(0) = \partial_y f(L) =0 , \\
			\partial_y \left( \frac{e^{-4W}}{\sqrt{g_{55}}} g \right)(0)=g(L)=0.
		\end{array}
	\right. 
\end{eqnarray}
It follows that the above boundary conditions ensure the hermiticity of the Hamiltonian, i.e.
\begin{equation}
	\langle \Psi_2 |H \Psi_1 \rangle =\langle H \Psi_2 | \Psi_1 \rangle.
\end{equation}
Therefore, we have succeeded to obtain the consistent set of boundary conditions that ensure the hermiticity of the supercharges and the Hamiltonian and also that the action of the supercharge on $|\Psi\rangle$ is well defined.
Since the supersymmetry is a direct consequence of higher-dimensional gauge invariance, our requirements on boundary conditions should be, at least, necessary conditions to preserve it.
It turns out that the boundary conditions obtained above are consistent with those in ref. \cite{Csaki}, although it is less obvious how the requirement of the least action principle proposed in ref. \cite{Csaki} is connected to gauge invariance. 
We should emphasize that the supercharge $Q$ is well defined for all the boundary conditions (4.16)-(4.19) and hence that the supersymmetric structure always appears in the spectrum, even though the boundary conditions other than the type (N,N) break 4d gauge symmetries, as we will see below.

From the above analysis, the 5d gauge fields $A_{\mu}(x,y)$ and $A_5(x,y)$ are expanded in the mass eigenstates as follows:
\begin{eqnarray}
	&& A_{\mu}(x,y) =\sum_n A_{\mu,n}(x)f_n(y) , \\
	&&A_5 (x,y)=\sum_n h_n(x)g_n(y) 
\end{eqnarray}
where $f_n(y)$ and $g_n(y)$ are the eigenstates of the Schr{\" o}dinger-like equations 
\begin{eqnarray}
	&&-\frac{1}{\sqrt{g_{55}}}\partial_y \frac{e^{-4W}}{\sqrt{g_{55}}}\partial_y f_n(y)=m_n^2 f_n(y) , \\
	&& -\partial_y \frac{1}{\sqrt{g_{55}}}\partial_y  \frac{e^{-4W}}{\sqrt{g_{55}}}g_n(y) = m_n^2 g_n(y) 
\end{eqnarray}
with one of the four types of the boundary conditions (4.16)-(4.19).
Since the massless states are especially important in phenomenology, let us investigate the massless states of the equations (4.23) and (4.24).
Thanks to supersymmetry, the massless modes would be the solutions to the first order differential equation $Q | \Psi_0 \rangle = 0$, i.e.
\begin{eqnarray}
	&& \partial_y f_0(y) = 0, \\
	&&\partial_y \left( \frac{e^{-4W}}{\sqrt{g_{55}}}g_0(y) \right)=0.
\end{eqnarray}
The solutions are easily found to be 
\begin{eqnarray}
	&&f_0(y)=c, \\
	&& g_0(y) =c' e^{4W(y)}\sqrt{g_{55}(y)},
\end{eqnarray}
where $c$ and $c'$ are some constants.
We should emphasize that the above solutions do not necessarily imply physical massless states of $A_{\mu,0}(x)$ and $h_0(x)$ in the spectrum.
This is because the boundary conditions exclude some or all of them from the physical spectrum.
Indeed, $f_0(y)$ $((g_0(y)))$ satisfies only the boundary conditions of the type (N,N) (type (D,D)).
Thus, a massless vector $A_{\mu,0}(x)$ (a massless scalar $h_0(x)$) appears only for the type (N,N) (type (D,D)) boundary conditions. 
This implies that the 4d gauge symmetry is broken except for the type (N,N) boundary conditions.

Let us next discuss geometrical meanings of the boundary conditions.
To this end, it is convenient to rewrite the equations (4.23) and (4.24) into a familiar form of the $N=2$ supersymmetric quantum mechanics.
Reparametrizing the coordinate $y$ such that
\begin{equation}
	ds^2 =e^{-4W(\tilde{y})}\left( \eta_{\mu \nu}dx^{\mu}dx^{\nu} +d\tilde{y}^2\right),
\end{equation}
where $W(\tilde{y})=W( y(\tilde{y}))$, we can rewrite the equations (4.23) and (4.24) into the form\footnote{
Shaposhnikov and Tinyakov\cite{Shaposhnikov-Tinyakov} have observed
that mass eigenfunctions for $A_{\mu}(x,y)$ satisfy a similar
equation to eq.(4.30) in a 5d gauge theory with a weight function.
The supersymmetric structure is, however, obscure due to the gauge
$A_5(x,y) = 0$.
}
\begin{eqnarray}
	&& -\bar{{\cal D}}_{\tilde{y}} {\cal D}_{\tilde{y}} \tilde{f}_n(\tilde{y})=m_n^2 \tilde{f}_n(\tilde{y}), \\
	&& -{\cal D}_{\tilde{y}} \bar{{\cal D}}_{\tilde{y}}\tilde{g}_n(\tilde{y})=m_n^2 \tilde{g}_n(\tilde{y}),
\end{eqnarray}
where
\begin{eqnarray}
	&&e^{W(\tilde{y})} \tilde{f}_n(\tilde{y}) =f_n(y) , \\
	&& e^{3W(\tilde{y})}\sqrt{g_{55}(\tilde{y})} \tilde{g}_n(\tilde{y}) =g_n(y) , \\
	&& {\cal D}_{\tilde{y}} =\partial_{\tilde{y}} + W'(\tilde{y}), \\
	&& \bar{{\cal D}}_{\tilde{y}}=\partial_{\tilde{y}} - W'(\tilde{y}).
\end{eqnarray}
Here, the prime denotes the derivative with respect to $\tilde{y}$.
The Hamiltonian and the supercharge can be written, in this basis, as
\begin{eqnarray}
	\tilde{H} &=&\tilde{Q}^2 =\left(	
		\begin{array}{cc}
			-\bar{{\cal D}}_{\tilde{y}} {\cal D}_{\tilde{y}}  & 0 \\
			0 & -{\cal D}_{\tilde{y}}\bar{{\cal D}}_{\tilde{y}}
		\end{array}
	\right) \nonumber \\
	&&\quad \  = \left(
		\begin{array}{cc}
			-\partial_{\tilde{y}}^2 -W''(\tilde{y})+\left(W'(\tilde{y})\right)^2 & 0 \\
			0&  -\partial_{\tilde{y}}^2 +W''(\tilde{y})+\left(W'(\tilde{y})\right)^2
		\end{array}
	\right), \\
	\tilde{Q}&=&\left(
		\begin{array}{cc}
			0 & -\bar{{\cal D}}_{\tilde{y}} \\
			{\cal D}_{\tilde{y}} &0 
		\end{array}
	\right).
\end{eqnarray}
These expressions are nothing but the $N=2$ supersymmetric quantum mechanics given by Witten\cite{Witten}, and $W(\tilde{y})$ is called a superpotential. 
In this basis, the boundary conditions (4.16)-(4.19) become
\begin{eqnarray}
	&&{\rm Type}\ ({\rm N,N}) : \left\{	
		\begin{array}{l}
			\tilde{f}'(\tilde{y}_0)+W'(\tilde{y}_0)\tilde{f}(\tilde{y}_0)=\tilde{f}'(\tilde{y}_L)+W'(\tilde{y}_L)\tilde{f}(\tilde{y}_L)=0 , \\
			\tilde{g}(\tilde{y}_0)=\tilde{g}(\tilde{y}_L)=0, 
		\end{array}
	\right. \\
	&&{\rm Type}\ ({\rm D,D}) : \left\{	
		\begin{array}{l}
			 \tilde{f}(\tilde{y}_0)=\tilde{f}(\tilde{y}_L)=0, \\
			\tilde{g}'(\tilde{y}_0)-W'(\tilde{y}_0)\tilde{g}(\tilde{y}_0)=\tilde{g}'(\tilde{y}_L)-W'(\tilde{y}_L)\tilde{g}(\tilde{y}_L)=0,
		\end{array}
	\right. \\
	&&{\rm Type}\ ({\rm N,D}) : \left\{	
		\begin{array}{l}
			\tilde{f}'(\tilde{y}_0)+W'(\tilde{y}_0)\tilde{f}(\tilde{y}_0)=\tilde{f}(\tilde{y}_L)=0, \\
			\tilde{g}(\tilde{y}_0)=\tilde{g}'(\tilde{y}_L)-W'(\tilde{y}_L)\tilde{g}(\tilde{y}_L)=0,
		\end{array}
	\right. \\
	&&{\rm Type}\ ({\rm D,N}) : \left\{	
		\begin{array}{l}
			\tilde{f}(\tilde{y}_0)=\tilde{f}'(\tilde{y}_L)+W'(\tilde{y}_L)\tilde{f}(\tilde{y}_L)=0, \\
			\tilde{g}'(\tilde{y}_0)-W'(\tilde{y}_0)\tilde{g}(\tilde{y}_0)=\tilde{g}(\tilde{y}_L)=0,
		\end{array}
	\right. 
\end{eqnarray}
where $\tilde{y}_0 =\tilde{y} (y=0)$ and $\tilde{y}_L =\tilde{y} (y=L)$.
For the Dirichlet boundary conditions of $\tilde{f}(\tilde{y})=0$ and $\tilde{g}(\tilde{y})=0$ at $\tilde{y}=\tilde{y}_0, \tilde{y}_L$, we can interpret them as the existence of rigid walls at the boundaries. 
For the other boundary conditions of $\tilde{f}'(\tilde{y})+W'(\tilde{y}) \tilde{f}(\tilde{y})=0$ and $\tilde{g}'(\tilde{y})-W'(\tilde{y})\tilde{g}(\tilde{y})=0$ at $\tilde{y}=\tilde{y}_0, \tilde{y}_L$, we can also interpret them as the existence of delta function potentials at the boundaries. 
Since for a delta function potential a localized (bound) state can appear, the low energy spectrum will have interesting properties for a non-trivial function $W(\tilde{y})$\cite{Nagasawa-Sakamoto,R-S}.

Before closing this section, it is instructive to investigate 5d gauge invariance in non-abelian gauge theories.
Let $G$ and $H$ be a non-abelian gauge group and its subgroup, respectively.
We consider the situation that the 4d gauge symmetry $G$ is broken to $H$ via boundary conditions. 
We denote the generators of $G, H$ and $G/H$ 
by $\{T^I\}, \{T^a\}$ and $\{T^{\hat{a}}\}$, respectively.
The 4d gauge symmetry breaking of $G \to H$ may be realized by imposing the type (N,N) boundary conditions on the gauge fields $A_M^a(x,y)$, which correspond to the unbroken generators of $H$, and one of the other boundary conditions on $A_M^{\hat{a}}(x,y)$, which correspond to the broken generators of $G/H$.

Infinitesimal gauge transformations will be given by 
\begin{equation}
	\delta A_M^I(x,y) = \partial_M \varepsilon^I(x,y) +g f^{IJK} A_M^J(x,y) \varepsilon^K(x,y), 
\end{equation}
where $g$ is the 5d gauge coupling constant and $f^{IJK}$ are the structure constants of $G$.
The boundary conditions for the gauge parameters $\varepsilon^I(x,y)$ should be taken to be the same as $A_{\mu}^I(x,y)$.
This requirement comes from the consistency with the inhomogeneous terms in eq. (4.42).
In order for the 5d gauge invariance under the transformations (4.42) to preserve, the homogeneous terms on the right-hand-side of eq. (4.42) have to obey the same boundary conditions as $A_M^I(x,y)$.
Then, it turns out that this is the case provided that\cite{BRS&Unitarity}
\begin{equation}
	f^{\hat{a}\hat{b}\hat{c}}=0.
\end{equation}
Although the above conditions are not, in general, satisfied for arbitrary choice of $G$ and $H$, they can be realized, for instance, if the $Z_2$ parity for the generators of $G$ are assigned as
\begin{eqnarray}
	&&{\cal P}(T^{a})= +T^a, \\
	&&{\cal P}(T^{\hat{a}})=-T^{\hat{a}}.
\end{eqnarray}
It is interesting to note that this happens for gauge symmetry breaking 
via $Z_2$-orbifolding\cite{OrbifoldGut}.
Thus, we have found that the 5d gauge invariance is preserved under the infinitesimal gauge transformations (4.42) with the conditions (4.43), even though the 4d gauge symmetry $G$ is broken to $H$ in the 4d effective theory.
\section{Summary and Discussions}

We have discovered the quantum-mechanical $N=2$ supersymmetry 
on any compact Riemannian manifolds without a boundary, 
and explicitly constructed the $N=2$ supercharges in the language of differential forms. 
The technology has been applied to gauge invariant theories
with extra dimensions, 
and the supersymmetric structure has been observed between 4d 
and extra-space components of gauge fields. 
The supersymmetry manifests itself in their 4d mass spectrum 
and massless 4d modes are found to be the solutions 
to the first order differential equation
\begin{equation}
	Q|\Psi_0 \rangle=0.
\end{equation}
It is then clear that the massless modes possess distinct analytic properties 
from other massive modes, which obey the 2nd order differential equations.

We have also discussed boundary conditions in gauge theories on extra dimensions with boundaries.
In a gauge symmetry point of view, it is less obvious to obtain a possible set of boundary conditions consistent with gauge invariance because some of the boundary conditions explicitly break 4d gauge symmetries.
On the other hand, in a supersymmetry point of view, the requirement of 5d gauge invariance is replaced by the conditions that the supercharges are hermitian and also that the action of the supercharges are well defined on a functional space with definite boundary conditions. 
In this framework, 4d gauge symmetry breaking via boundary conditions\footnote{Here, we say that a 4d gauge symmetry is broken if there is no massless 4d gauge boson corresponding to the gauge symmetry.}
may be interpreted as \lq\lq spontaneous" supersymmetry breaking with no zero energy state\footnote{We use the words, \lq\lq spontaneous" supersymmetry breaking, by analogy with supersymmetric quantum field theory, in which if there is no zero energy state, supersymmetry is spontaneously broken.
Then, supercharges become ill defined and the degeneracy between bosons and fermions is lost.}.
We should emphasize that the supercharges are well defined in quantum mechanics even if there is no zero energy state, and hence that the degeneracy between \lq\lq bosonic" and \lq\lq fermionic" states still holds.
In this sense, the boundary conditions we obtained are consistent with 5d gauge invariance, even if some of 4d gauge symmetries are broken via boundary conditions.

Since the origin of the supersymmetry is the gauge invariance in higher dimensions, we expect that any higher dimensional theories with gauge-like symmetries possess supersymmetry.
Such an example is a gauge theory with an antisymmetric field, which often appear in string theory.
Since the action of the antisymmetric gauge field can be written in terms of differential forms, it will be straightforward to show the $N=2$ supersymmetric structure of the theory.
It turns out that the $N=2$ supersymmetry is actually enhanced 
in particular dimensions and that the $N=2$ supersymmetry algebra given in Section 2 
can be extended to an $N=4$ supersymmetry algebra by adding a duality operator.
The results will be reported elsewhere\cite{Preparation}.
\renewcommand{\theequation}{A.\arabic{equation}}
\appendix
\section{HODGE DECOMPOSITION THEOREM}

In this appendix, we give a simple proof of the Hodge decomposition theorem by use of the eigenfunctions of the differential operators $d^{\dagger}d$ and $d d^{\dagger}$.

Since $d^{\dagger}d$ is a hermitian operator, the eigenfunctions will form a complete set.
Thus, any $k$ form $A^{(k)}$ can be expanded as
\begin{equation}	A^{(k)}=\omega_0^{(k)}+\textstyle{\displaystyle{\sum_{n_k}}'} a_{n_k}\omega_{n_k}^{(k)},
\end{equation}
where
\begin{eqnarray}
	d\omega_0^{(k)}&=&0 , \\
	d^{\dagger}d \omega_0^{(k)}&=&(m_{n_k}^{(k)})^2\omega_{n_k}^{(k)} \qquad {\rm for}\ m_{n_k}^{(k)}\ne 0 .
\end{eqnarray}
Here, $\textstyle{\displaystyle{\sum_{n_k}}'}$ denotes the summation over all eigenstates with $m_{n_k}^{(k)} \ne 0$.
Since $d d^{\dagger}$ is also a hermitian operator, $\omega_0^{(k)}$ can be expanded as 
\begin{equation}
	\omega_0^{(k)} =\eta_0^{(k)}+\textstyle{\displaystyle{\sum_{n_{k-1}}}'} b_{n_{k-1}}\phi_{n_{k-1}}^{(k)},
\end{equation}
where 
\begin{eqnarray}
	&&d^{\dagger}\eta_0^{(k)}=d \eta_0^{(k)}=0 , \\
	&& dd^{\dagger}\phi_{n_{k-1}}^{(k)}=( m_{n_{k-1}}^{(k-1)})^2\phi_{n_{k-1}}^{(k)} \qquad {\rm for} \ m_{n_{k-1}}^{(k-1)}\ne 0.
\end{eqnarray}
It is convenient to further introduce the eigenfunctions of $dd^{\dagger}$ for $k+1$ forms and $d^{\dagger}d$ for $k-1$ forms as
\begin{eqnarray}
	dd^{\dagger}\phi_{n_k}^{(k+1)}&=&(m_{n_k}^{(k)})^2\phi_{n_k}^{(k+1)}, \\
	 d^{\dagger}d \omega_{n_{k-1}}^{(k-1)}&=&(m_{n_{k-1}}^{(k-1)})^2\omega_{n_{k-1}}^{(k-1)}.
\end{eqnarray}
As shown in Section 2, $\omega_{n_k}^{(k)}$ and $\phi_{n_{k-1}}^{(k)}$ are related to $\phi_{n_k}^{(k+1)}$ and $\omega_{n_{k-1}}^{(k-1)}$ as 
\begin{eqnarray}
	m_{n_k}^{(k)}\omega_{n_k}^{(k)}=d^{\dagger}\phi_{n_k}^{(k+1)} \quad &{\rm or}& \quad m_{n_k}^{(k)}\phi_{n_k}^{(k+1)}=d\omega_{n_k}^{(k)}, \\
	m_{n_{k-1}}^{(k-1)}\omega_{n_{k-1}}^{(k-1)}=d^{\dagger}\phi_{n_{k-1}}^{(k)} \quad &{\rm or}& \quad m_{n_{k-1}}^{(k-1)}\phi_{n_{k-1}}^{(k)}=d\omega_{n_{k-1}}^{(k-1)}.
\end{eqnarray}
A $k$ form satisfying eqs.(A.5) is called a harmonic $k$ form which can be expanded, in terms of a complete set of the harmonic $k$ forms $\{\eta_p^{(k)}, \ p=1,2,\cdots, b_k\}$, as 
\begin{equation}
	\eta_0^{(k)} =\sum_{p=1}^{b_k}c_p \eta_p^{(k)}.
\end{equation}
The integer $b_k$, which is the number of the independent harmonic $k$ forms, is called the $k$th Betti number and is known as a topological number of the manifold.

We have thus found that any $k$ form $A^{(k)}$ can be expanded as 
\begin{eqnarray}
	A^{(k)}&=&\sum_{p=1}^{b_k} c_p \eta_p^{(k)} +\textstyle{\displaystyle{\sum_{n_k}}'} a_{n_k}\omega_{n_k}^{(k)}+\textstyle{\displaystyle{\sum_{n_{k-1}}}'}b_{n_{k-1}}\phi_{n_{k-1}}^{(k)} \nonumber \\
	&=& \sum_{p=1}^{b_k}c_p \eta_p^{(k)}+\textstyle{\displaystyle{\sum_{n_k}}'}\displaystyle{\frac{a_{n_k}}{m_{n_k}^{(k)}}}d^{\dagger}\phi_{n_k}^{(k+1)}+\textstyle{\displaystyle{\sum_{n_{k-1}}}'}\displaystyle{\frac{b_{n_{k-1}}}{m_{n_{k-1}}^{(k-1)}}}d\omega_{n_{k-1}}^{(k-1)}.
\end{eqnarray}
This implies that any $k$ form has the decomposition of the form 
\begin{equation}
	A^{(k)}=\eta_0^{(k)}+d^{\dagger}\alpha^{(k+1)}+d\beta^{(k-1)}.
\end{equation}
This completes a proof of the Hodge decomposition theorem.
\section*{Acknowledgements}
C.S.L., H.S. and M.S. are supported in part by the Grant-in-Aid for
Scientific Research (No.15340078, No.14340077 and No.15540277) by the Japanese Ministry
of Education, Science, Sports and Culture.
\baselineskip 5mm 


\begin{thebibliography}{99}
\bibitem{SS}
J. Scherk and J. H. Shwarz, Nucl. Phys. {\bf B153} (1979), 61; Phys. Lett. {\bf B82} (1979), 60.\\
 P. Fayet, Phys. Lett. {\bf B159} (1985), 121; Nucl. Phys. {\bf B263} (1986), 649. 
\bibitem{hosotani}
	Y. Hosotani, Phys. Lett. {\bf B126} (1983), 309; Ann. Phys. {\bf 190} (1989), 233. 
\bibitem{G-H}
	H. Hatanaka, T. Inami and C. S. Lim, Mod. Phys. Lett. {\bf A13} (1998), 2601, hep-th/9805067.\\
   M. Kubo, C. S. Lim and H. Yamashita, Mod. Phys. Lett. {\bf A17} (2002), 2249, hep-ph/0111327. \\
   L. J. Hall, Y. Nomura and D. R. Smith, Nucl. Phys. {\bf B639} (2002), 307, hep-ph/0107331. \\
   G. Burdman and Y. Nomura, Nucl. Phys. {\bf B656} (2003), 3, hep-ph/0210257. \\
  N. Haba, Y. Hosotani, Y. Kawamura and T. Yamashita, Phys. Rev. {\bf D70} (2004), 015010, hep-ph/0401183. \\
  N. Haba, K. Takenaga and T. Yamashita, hep-ph/0411250. \\
  Y. Hosotani, S. Noda and K. Takenaga, Phys. Rev. {\bf D69} (2004), 125014, hep-ph/0403106; hep-ph/0410193.
 \bibitem{OrbifoldGut}	
	Y. Kawamura, Prog. Theor. Phys. {\bf 103} (2000), 613, hep-ph/9902423; Prog. Theor. Phys. {\bf 105} (2001), 691, hep-ph/0012352; Prog. Theor. Phys. {\bf 105} (2001), 999, hep-ph/0012125. \\
	L. J. Hall and Y. Nomura, Phys. Rev. {\bf D64} (2001), 055003, hep-ph/0103125. \\
	A. Hebecker and J. March-Russell, Nucl. Phys. {\bf B613} (2001), 3, hep-ph/0106166; Nucl. Phys. {\bf B625} (2002), 128, hep-ph/0107039. 
\bibitem{Nagasawa-Sakamoto}
	T. Nagasawa and M. Sakamoto, Prog. Theor. Phys. {\bf 112} (2004), 629, hep-ph/0406024; hep-ph/0410383.
\bibitem{Csaki}
   C. Cs{\' a}ki, C. Grojean, H. Murayama, L. Pilo, J. Terning, Phys. Rev. {\bf D69} (2004), 055006, hep-ph/0305237. 
\bibitem{Higgsless}
   C. Cs{\' a}ki, C. Grojean, L. Pilo, J. Terning, Phys. Rev. Lett. {\bf 92} (2004), 101802, hep-ph/0308038. \\
   Y. Nomura, JHEP {\bf 0311} (2003), 050, hep-ph/0309189. \\
   S. Gabriel, S. Nandi and G. Seidl, Phys. Lett. {\bf B603} (2004), 74, hep-ph/0406020. \\
   C. Schwinn, Phys. Rev. {\bf D69} (2004), 116005, hep-ph/0402118. \\
   G. Cacciapaglia, C. Cs{\' a}ki, C. Grojean and J. Terning, Phys. Rev. {\bf D70}  (2004), 075014, hep-ph/0401160.  \\
   G. Burdman and Y. Nomura, Phys. Rev. {\bf D69} (2004), 115013, hep-ph/0312247.  \\
   C. Cs{\' a}ki, hep-ph/0412339. 
\bibitem{arkani-cohen-georgi} 
    N. Arkani-Hamed, A. G. Cohen and H. Georgi, Phys. Lett. {\bf B513} (2001), 232,
    hep-ph/0105239.
\bibitem{deconstruction}
	R. S. Chivukula, E. H. Simmons, H. J. He, M. Kurachi and M. Tanabashi, Phys. Rev. {\bf D70} (2004), 075008, hep-ph/0406077. \\
	H. Georgi, hep-ph/0408067. \\
	R. Foadi, S. Gopalakrishna and C. Schmidt, JHEP {\bf 0403} (2004), 042,
    hep-ph/0312324. 
\bibitem{BRS&Unitarity}
 	T. Ohl, C. Schwinn, Phys. Rev. {\bf D70} (2004), 045019, hep-ph/0312263. \\
 	Y. Abe, N. Haba, K. Hayakawa, Y. Matsumoto, M. Matsunaga and K. Miyachi, 
	hep-th/0402146. 
\bibitem{R-S}
    L. Randall and R. Sundrum, Phys. Rev. Lett. {\bf 83} (1999), 3370, hep-ph/9905221; Phys. Rev. Lett. {\bf 83} (1999), 4690, hep-th/9906064. 
\bibitem{Nagasawa-Sakamoto-Takenaga}
	 T. Nagasawa, M. Sakamoto, K. Takenaga, Phys. Lett. {\bf B562} (2003), 358, hep-th/0212192; Phys. Lett. {\bf B583} (2004), 357, hep-th/0311043. 
\bibitem{Shaposhnikov-Tinyakov}
    M. Shaposhnikov and P. Tinyakov, Phys. Lett. {\bf B515} (2001), 442,
    hep-th/0102161.	 
\bibitem{Witten}
    E. Witten, Nucl. Phys. {\bf B188} (1981), 513.
 \bibitem{Preparation}
 C. S. Lim, T. Nagasawa, M. Sakamoto, H. Sonoda, in preparation. 
 \end{thebibliography}
\end{document}